%%%%%%%%%%%%%%%%%%%%%%%%%%%%%%%%%%%%%%%%%%%%%%%%%%%%%%%%%%%%%%%%%%%%%%
%%%%%%%%%%%%%%%%%%%%%%%%%%%%%%%%%%%%%%%%%%%%%%%%%%%%%%%%%%%%%%%%%%%%%%
%%%                                                                %%%
%%%         .         .           ..           .         .         %%%
%%%             On the uselessness of quantum queries              %%%
%%%                                                                %%%
%%%             David A. Meyer and James Pommersheim               %%%
%%%                                                                %%%
%%%                                                                %%%
%%%%%%%%%%%%%%%%%%%%%%%%%%%%%%%%%%%%%%%%%%%%%%%%%%%%%%%%%%%%%%%%%%%%%%
%%%%%%%%%%%%%%%%%%%%%%%%%%%%%%%%%%%%%%%%%%%%%%%%%%%%%%%%%%%%%%%%%%%%%%
%%%                                                                %%%
%%%                                                                %%%
%%%                       Typesetting notes                        %%%
%%%                                                                %%%
%%%                                                                %%%
%%% Plain TeX with a few simple macros at the beginning.           %%%
%%%                                                                %%%
%%% If AMS fonts are not loaded, comment the next line and         %%%
%%%  uncomment the following one.                                  %%%
\font\bbb=msbm10                                    %%%
%\def\bbb{\bf} \def\bbs{\bf}                                       %%%
%%%                                                                %%%
%%%                                                                %%%
%%%%%%%%%%%%%%%%%%%%%%%%%%%%%%%%%%%%%%%%%%%%%%%%%%%%%%%%%%%%%%%%%%%%%%
%%%%%%%%%%%%%%%%%%%%%%%%%%%%%%%%%%%%%%%%%%%%%%%%%%%%%%%%%%%%%%%%%%%%%%

\overfullrule=0pt

\input epsf.tex
\input pstricks  
\input color
\definecolor{Green}{rgb}{0.25,0.75,0.25}
\definecolor{review}{rgb}{0.25,0.75,0.25}
\definecolor{exercise}{rgb}{1,0,0}
\definecolor{lightgray}{rgb}{0.5,0.5,0.5}

\def\C{\hbox{\bbb C}}  
\def\N{\hbox{\bbb N}}  
  
\def\Z{\hbox{\bbb Z}}  
\def\Tr{{\rm Tr}}
\def\Pr{{\rm Pr}}

\def\CACM{{\sl Commun.\ ACM\/}}

\def\JACM{{\sl J. ACM\/}}

\def\ML{{\sl Machine Learning}}

\def\PRL{{\sl Phys.\ Rev.\ Lett.}}
\def\PRSLA{{\sl Proc.\ Roy.\ Soc.\ Lond.\ A\/}}

\def\QIP{{\sl Quantum Inform.\ Processing}}

\def\SIAMJC{{\sl SIAM J. Comput.}}

\def\dajm{\hbox{D. A. Meyer}}

\def\jamie{\hbox{J. Pommersheim}}
\def\dj{\hbox{\dajm\ and \jamie}}

\def\grover{\hbox{L. K. Grover}}

\def\shor{\hbox{P. W. Shor}}
\def\simon{\hbox{D. R. Simon}}

\def\hfb{\hfil\break}

\catcode`@=11
\newskip\ttglue

   \font\ninerm=cmr9    \font\eightrm=cmr8   \font\sixrm=cmr6
  \font\ninebf=cmbx9   \font\eightbf=cmbx8  \font\sixbf=cmbx6
  \font\nineit=cmti9   \font\eightit=cmti8  
  \font\ninesl=cmsl9   \font\eightsl=cmsl8  
  \font\ninemi=cmmi9   \font\eightmi=cmmi8  \font\sixmi=cmmi6

\font\bigten=cmr10 scaled\magstep2 

\def\ninepoint{\def\rm{\fam0\ninerm}%
  \textfont0=\ninerm \scriptfont0=\sixrm
  \textfont1=\ninemi \scriptfont1=\sixmi
  \textfont\itfam=\nineit  \def\it{\fam\itfam\nineit}%
  \textfont\slfam=\ninesl  \def\sl{\fam\slfam\ninesl}%
  \textfont\bffam=\ninebf  \scriptfont\bffam=\sixbf
    \def\bf{\fam\bffam\ninebf}%
  \tt \ttglue=.5em plus.25em minus.15em
  \normalbaselineskip=11pt
  \setbox\strutbox=\hbox{\vrule height8pt depth3pt width0pt}%
  \normalbaselines\rm}

\def\eightpoint{\def\rm{\fam0\eightrm}%
  \textfont0=\eightrm \scriptfont0=\sixrm
  \textfont1=\eightmi \scriptfont1=\sixmi
  \textfont\itfam=\eightit  \def\it{\fam\itfam\eightit}%
  \textfont\slfam=\eightsl  \def\sl{\fam\slfam\eightsl}%
  \textfont\bffam=\eightbf  \scriptfont\bffam=\sixbf
    \def\bf{\fam\bffam\eightbf}%
  \tt \ttglue=.5em plus.25em minus.15em
  \normalbaselineskip=9pt
  \setbox\strutbox=\hbox{\vrule height7pt depth2pt width0pt}%
  \normalbaselines\rm}

\def\sfootnote#1{\edef\@sf{\spacefactor\the\spacefactor}#1\@sf
      \insert\footins\bgroup\eightpoint
      \interlinepenalty100 \let\par=\endgraf
        \leftskip=0pt \rightskip=0pt
        \splittopskip=10pt plus 1pt minus 1pt \floatingpenalty=20000
        \parskip=0pt\smallskip\item{#1}\bgroup\strut\aftergroup\@foot\let\next}
\skip\footins=12pt plus 2pt minus 2pt
\dimen\footins=30pc

\def\ie{{\it i.e.}}
\def\eg{{\it e.g.}}
\def\etc{{\it etc.}}
\def\etal{{\it et al.}}

\def\calc{{\cal C}}

\def\calh{{\cal H}}
\def\calof{{\cal O}_{\!f}}
\def\uhat{\hat{u}}
\def\qhat{\hat{q}}
\def\sign{{\rm sign}}

\def\Lemma{L{\eightpoint EMMA}}

\def\Theorem{T{\eightpoint HEOREM}}
\def\Example{E{\eightpoint XAMPLE}}
\def\Corollary{C{\eightpoint OROLLARY}}
\def\Definition{D{\eightpoint EFINITION}}

\def\endproof{\vrule height6pt width4pt depth2pt}

\def\and{{\eightpoint AND}}

\def\Simon{1}
\def\Shor{2}
\def\BonehLipton{3}
\def\Jozsa{4}
\def\Angluinsurvey{5}
\def\Grover{6}
\def\Deutsch{7}
\def\CEMM{8}
\def\BBCMdW{9}
\def\FGGS{10}
\def\Shamir{11}
\def\BVdW{12}
\def\MeyerPommersheim{13}
\def\MNR{14}
\def\MeyerPommersheimB{15}
\def\HMPPR{16}
\def\Angluin{17}
\def\BrassardHoyer{18}
\def\GroverB{19}
\def\BHT{20}
\def\BHMT{21}

\magnification=1200

\parskip=0pt\parindent=0pt

\line{\hfil 25 March 2010}
%\line{\hfill quant-ph/yymmnnn}
\vfill
\centerline{\bf\bigten ON THE USELESSNESS OF QUANTUM QUERIES}
\bigskip\bigskip
\centerline{\bf David A. Meyer$^*$ and 
                James Pommersheim$^{*,\dagger}$}
\bigskip 
\centerline{\sl $^*$Project in Geometry and Physics,
                Department of Mathematics}
\centerline{\sl University of California/San Diego,
                La Jolla, CA 92093-0112}
\smallskip
\centerline{\sl $^{\dagger}$Department of Mathematics}
\centerline{\sl Reed College, Portland, OR 97202-8199}
\smallskip
\centerline{{\tt dmeyer@math.ucsd.edu},
            {\tt jamie@reed.edu}}            
\smallskip

\vfill
\centerline{ABSTRACT}
\bigskip
%--------|---------|---------|---------|---------|---------|---------|
\noindent Given a prior probability distribution over a set of 
possible oracle functions, we define a number of queries to be
{\sl useless\/} for determining some property of the function if the 
probability that the function has the property is unchanged after the
oracle responds to the queries.  A familiar example is the parity of a 
uniformly random Boolean-valued function over $\{1,2,\ldots,N\}$, for 
which $N-1$ classical queries are useless.  We prove that if $2k$ 
classical queries are useless for some oracle problem, then $k$ 
quantum queries are also useless.  For such problems, which include 
classical threshold secret sharing schemes, our result also gives a 
new way to obtain a lower bound on the quantum query complexity, even 
in cases where neither the function nor the property to be determined 
is Boolean.

\bigskip\bigskip
%--------|---------|---------|---------|---------|---------|---------|
\noindent 2010 Physics and Astronomy Classification Scheme:
                   03.67.Ac. % Quantum algorithms, protocols, and
                             %  simulations

\noindent 2010 American Mathematical Society Subject Classification:
                   68Q12,    % Computer science
                             %  Theory of computing 
                             %   Quantum algorithms and complexity
                   68Q17,    % Computer science
                             %  Theory of computing
                             %   Computational difficulty of problems 
                             %   (lower bounds, completeness, 
                             %   difficulty of approximation, etc.)
                   81P68.    % Quantum theory
                             %  Axiomatics, foundations, philosophy 
                             %   Quantum computation

\smallskip
\global\setbox1=\hbox{Key Words:\enspace}
\parindent=\wd1
\item{Key Words:}  computational learning theory, oracle problem, 
                   query complexity, lower bound.

\vfill
\eject

\headline{\ninepoint\it On the uselessness of quantum queries 
                                         \hfill  Meyer \& Pommersheim}

\parskip=10pt
\parindent=0pt

\noindent{\bf 1.  Introduction}

%--------|---------|---------|---------|---------|---------|---------|
Many computational problems involve queries to an oracle (calls to a 
subroutine) that evaluates some function $f$ at the argument $x$
passed to it and returns the result $f(x)$.  Typically, the task is to 
use the oracle to determine some property of the unknown function.  An 
important example for quantum computation is P{\eightpoint ERIOD} 
F{\eightpoint INDING} [\Simon,\Shor] (the A{\eightpoint BELIAN} 
H{\eightpoint IDDEN} S{\eightpoint UBGROUP} P{\eightpoint ROBLEM} 
[\BonehLipton,\Jozsa]), where the function is invariant under addition 
of some constant to its argument and the task is to find that 
constant.  Another example is C{\eightpoint ONCEPT} 
L{\eightpoint EARNING}, where there is some set (the {\sl concept 
class\/}) of Boolean-valued functions and the task is to identify 
exactly which one (the {\sl concept\/}) the oracle is evaluating
[\Angluinsurvey].  Grover's U{\eightpoint NSTRUCTURED} 
S{\eightpoint EARCH} problem [\Grover] is an instance of concept 
learning, where the possible functions each take the value $1$ for 
exactly one argument and the value $0$ for all other arguments.

%--------|---------|---------|---------|---------|---------|---------|
A natural goal is to minimize the number of queries to the oracle 
needed to solve the problem; this minimum is the {\sl query 
complexity\/} of the problem.  An alternative goal is to maximize the 
probability of determining the desired property of $f$ using no more 
than some fixed number of queries, $k$.  Although this probability is 
clearly non-decreasing in $k$, when it does not increase with 
additional queries, we might say that these queries {\sl provide no 
information}, or describe them as {\sl useless}.

%--------|---------|---------|---------|---------|---------|---------|
For example, consider Deutsch's problem, in which
$f : \{1,2\}\to\Z_2$ and the property to be determined is $f(1)+f(2)$ 
[\Deutsch].  If $f$ is chosen uniformly at random, then the prior 
probabilities for the value of this sum are each $1/2$.  In this case, 
a single classical query is useless:  the posterior probabilities for 
the value of the sum are unchanged after the oracle responds to either 
query.  A single quantum query, on the other hand, is not useless:  
used properly, it identifies the value of the sum with probability $1$ 
[\CEMM].

%--------|---------|---------|---------|---------|---------|---------|
But this raises a natural question:  Can quantum queries be useless?
In this paper, we formalize the notion of uselessness and study 
problems for which the answer to this question is ``yes''.  Our main
result is a relation between the uselessness of classical and quantum 
queries:  if $2k$ classical queries provide no relevant information 
about $f$, then $k$ quantum queries provide no relevant information 
about $f$.

%--------|---------|---------|---------|---------|---------|---------|
The maximum number of queries that is useless will always be a lower 
bound for the query complexity; thus our analysis provides a new 
method for finding a lower bound for the quantum query complexity of
any problem for which some number of classical queries is useless.

%--------|---------|---------|---------|---------|---------|---------|
A familiar problem to which our results apply is P{\eightpoint ARITY}, 
a generalization of Deutsch's problem in which $N\in\N$ is fixed, an 
arbitrary function $f:\{1,\ldots,N\}\to\Z_2$ is chosen uniformly at 
random, and the property to be determined is the modulo $2$ sum of the 
values of $f$.  This problem is an example of a {\sl black box\/} 
oracle problem, in which the values $f(1),\ldots,f(N)$ form an unknown 
$N$-bit string.  Since $N-1$ classical queries reveal no information 
about the parity of this string, our result says that 
$\lfloor(N-1)/2\rfloor$ quantum queries are also useless.  This 
implies that the quantum query complexity must be at least 
$\lceil N/2\rceil$.  Beals, \etal\ obtain this same lower bound using 
the polynomial method, and note that this bound is realized by a 
quantum algorithm that applies the solution to Deutsch's problem to 
the function values in pairs [\BBCMdW].%
\parindent=9pt
\sfootnote{$^*$}{Farhi, \etal\ obtained the same results using a
different method [\FGGS].}
\parindent=0pt%

%--------|---------|---------|---------|---------|---------|---------|
P{\eightpoint ARITY} is a simple example of a classical {\sl threshold 
secret sharing scheme}.  In the oracle problem framework, a classical 
$(k,N)$ threshold secret sharing scheme [\Shamir] can be described as 
a set of functions $f:\{1,\ldots,N\}\to Y$ together with some property 
of $f$ that can be determined by any $k$ distinct classical queries, 
but about which no $k-1$ classical queries provide any information.  
So as a corollary of our main theorem, we find that any classical 
$(k,N)$ threshold secret sharing scheme defines an oracle problem for 
which $\lfloor(k-1)/2\rfloor$ quantum queries are useless and which 
therefore has quantum query complexity at least $\lceil k/2\rceil$.

%--------|---------|---------|---------|---------|---------|---------|
Thus our results also give new quantum lower bounds, \eg, for 
P{\eightpoint OLYNOMIAL} I{\eightpoint NTERPOLA\-TION}, a threshold 
secret sharing scheme introduced by Shamir [\Shamir].  Here the 
function $f$ is a polynomial function of degree $k$ over $\Z_p$, with 
$k+1<p$, and the problem is to determine $f(0)$.  The theory of 
polynomials easily implies that $k+1$ classical queries suffice, but 
$k$ queries yield no information.  Applying our general results, this
implies that $\lfloor k/2\rfloor$ quantum queries yield no 
information and thus at least $\lfloor k/2\rfloor + 1$ quantum queries 
are necessary.

%--------|---------|---------|---------|---------|---------|---------|
As this problem exemplifies, our formulation includes oracles that 
return more than a single bit in response to a query; it is thus more 
general than the one in which query complexities of Boolean functions 
are studied.  Moreover, as indicated in the discussion of Deutsch's 
problem above, our formulation also includes a prior probability 
distribution over possible oracles.  As such it includes the more 
commonly studied cases of total and partial functions as special 
cases:  the former has a constant probability distribution over all 
functions, while the latter has a two-valued probability distribution 
that vanishes on disallowed functions. 

%--------|---------|---------|---------|---------|---------|---------|
Furthermore, the methods we use to prove the main theorem are new.  In 
the Appendix, we show how an existing method, the polynomial lower 
bound method [\BBCMdW], together with an observation of Buhrman, 
\etal~[\BVdW], can be used to prove a special case of our theorem, 
namely the case in which we wish to compute a Boolean function, or 
partial Boolean function, of an $N$-bit string.  But these existing 
methods do not appear to suffice to prove our theorem in complete 
generality, \ie, in their current form they do not apply to the case 
in which the set $Y$ has more than $2$ elements, nor to the case in 
which we wish to compute more than just a Boolean classification of 
the allowed functions.

\medskip
\noindent{\bf 2.  The definition of uselessness}

\nobreak
%--------|---------|---------|---------|---------|---------|---------|
Let $X$ and $Y$ be finite sets, and let $\calc\subseteq Y^X$ be a 
subset of the set of all functions from $X$ to $Y$.  Boolean-valued 
functions, \ie, $Y = \Z_2$, are commonly studied---in computational 
learning theory, for example, where $\calc$ is called a 
{\sl concept class\/} [\Angluinsurvey].

%--------|---------|---------|---------|---------|---------|---------|
Suppose that the class $\calc$ is partitioned into disjoint subclasses 
$\calc_j$, $j\in J$.  In the {\sl learning problems\/} 
$\bigl(\calc,\{\calc_j\mid j\in J\},\mu\bigr)$ we are 
considering, an element $f$ is chosen from $\calc$ according to an 
arbitrary, but known, prior probability distribution $\mu$, and the 
task is to determine to which subclass $\calc_j$ the function $f$ 
belongs.  Information about $f$ is available only {\it via\/} an 
oracle that, given a query $x\in X$, returns the value of $f(x)$.  To 
formalize the action of this oracle we begin by recalling some 
standard notation:

%--------|---------|---------|---------|---------|---------|---------|
Let $\calh = \C^X \otimes \C^Y \otimes \C^Z $, where $Z$ is a finite 
set.  The three tensor factors represent query, response, and 
auxiliary registers, respectively.  We assume that $Y$ is an abelian 
group, and that the quantum oracle $\calof$ acts on $\calh$ by 
addition of $f(x)$ into the response register.  (Everything in the
following, however, can be carried out more generally in the 
permutation model introduced in [\MeyerPommersheim].)  Thus the action 
of the oracle $\calof : \calh\to\calh$ is specified by the following 
permutation of the computational basis
$$
\calof: |x,y,z\rangle \mapsto |x,y+f(x),z\rangle.
$$

%--------|---------|---------|---------|---------|---------|---------|
A general $k$-query quantum learning algorithm can now be described as 
follows:  An initial state is prepared with density matrix 
$\rho_0\in \calh\otimes\calh^{\dagger}$.  The algorithm passes this
state to the oracle, which acts by $\calof$; then the algorithm acts 
by some unitary operator $U_1$, independent of $f$; and the state is 
again passed to the oracle; \etc\ \ After the $k^{\rm th}$ call to the 
oracle, the algorithm applies a last unitary operator $U_k$ to arrive 
in the final state
$$
\rho_f = 
U_k^{\vphantom\dagger} 
 \calof^{\vphantom\dagger} 
  U_{k-1}^{\vphantom\dagger} 
   \ldots 
    U_1^{\vphantom\dagger} 
     \calof^{\vphantom\dagger} 
      \rho_0 
     \calof^{\dagger}
    U_1^{\dagger}
   \ldots 
  U_{k-1}^{\dagger}
 \calof^{\dagger}
U_k^{\dagger}.                                                \eqno(1)
$$
The last step is a POVM $\{\Pi_s\}$ indexed by an arbitrary set $S$.
Some map $S\to J$, which is part of the algorithm (and independent of
$f$), specifies the subset $\calc_j$ to which we conclude $f$ belongs.  
(Notice that the unitary operator $U_k$ is unnecessary, since it could 
be incorporated into the measurement.  It is notationally convenient, 
however, to include it.)

%--------|---------|---------|---------|---------|---------|---------|
Our main result concerns situations in which no information about the
part $\calc_j$ to which the function $f$ belongs can be derived from 
some number of classical or quantum queries.  We now make this notion 
precise.

%--------|---------|---------|---------|---------|---------|---------|
\noindent\Definition\ (classical version).  Let 
$\bigl(\calc,\{\calc_j\mid j\in J\},\mu\bigr)$ be a learning problem 
as described above.  Then we say that $k$ classical queries 
{\sl yield no information}, or are {\sl useless}, if for any
$x_1,\ldots,x_k\in X$ and $y_1,\ldots,y_k\in Y$, 
$$
\mu\bigl(f\in\calc_j\mid f(x_i)= y_i, i=1,\ldots,k\bigr) 
 = \mu(f\in\calc_j),
\hbox{\ for\ all\ $j\in J$}.  
$$
That is, the probability of $f$ being in any of the sets $\calc_j$ is 
independent of the knowledge of any $k$ function values.

%--------|---------|---------|---------|---------|---------|---------|
\noindent\Definition\ (quantum version).   Let 
$\bigl(\calc,\{\calc_j\mid j\in J\},\mu\bigr)$ be a learning problem 
as described above.  Then we say that $k$ quantum queries 
{\sl yield no information}, or are {\sl useless}, if for any $k$-query 
quantum algorithm with initial state $\rho_0$, unitary operations 
$U_1,\ldots,U_k$, and measurement $\{\Pi_s\}$, 
$$
\mu(f\in\calc_j\mid s) = \mu(f\in\calc_j),
\hbox{\ for\ all\ $s\in S$,\ $j\in J$.}
$$
That is, the probability of $f$ being in $\calc_j$ is independent of 
any measurement taken after $k$ calls to the oracle.

\medskip
\noindent{\bf 3.  From classical to quantum uselessness}

%--------|---------|---------|---------|---------|---------|---------|
Having made these definitions precise, we can state our main result:

%--------|---------|---------|---------|---------|---------|---------|
\noindent\Theorem~1.  {\sl Let 
$\bigl(\calc,\{\calc_j\mid j\in J\},\mu\bigr)$ be a learning problem.  
Suppose that $2k$ classical queries are useless.  Then $k$ quantum 
queries are useless.}

%--------|---------|---------|---------|---------|---------|---------|
\noindent\Example~1 (P{\eightpoint ARITY}).  As we described in the
introduction, Theorem~1 applies to P{\eightpoint ARITY}:  Let 
$N\in\N$, and let $\calc$ be the set of all functions from 
$\{1,\ldots,N\}$ to $\Z_2$ with a uniform prior distribution.  
Partition $\calc$ into $\calc_0$ and $\calc_1$ according to the sum of 
the values of $f$.  Then it is easy to see that $N-1$ classical 
queries are useless.  Thus, by Theorem 1, $\lfloor{(N-1)/2}\rfloor$ 
quantum queries are also useless.  Since P{\eightpoint ARITY} can be 
solved with $\lceil{N/2}\rceil$ quantum queries (using repeated 
{\eightpoint XOR}s, \ie, solutions to Deutsch's problem), the quantum 
query complexity of P{\eightpoint ARITY} for exact solution is exactly 
$\lceil{N/2}\rceil$, reproving a result of Farhi, \etal\ [\FGGS] and 
Beals, \etal\ [\BBCMdW].  Theorem~1 tells us a little more, namely 
that using $1$ fewer query than this there is no quantum algorithm 
that succeeds with probability greater than $1/2$, a result that we
show in the Appendix also follows from the analysis of {\sl unbounded 
error quantum query complexity\/} of Boolean functions by Montanaro, 
\etal\ using more complicated machinery [\MNR].

%--------|---------|---------|---------|---------|---------|---------|
\noindent\Example~2.  Generalizing Deutsch's problem in a different 
direction than does P{\eightpoint ARITY}, let $\calc$ be the set of 
all functions from $\{1,2,3\}$ to $\Z_3$ with a uniform prior 
distribution.  Let 
$\calc = \calc_{{\rm even}}\sqcup\calc_{{\rm odd}}$, where a function 
$f$ is defined to be even or odd depending on whether the size of the 
image of $f$ is even or odd.  Notice that the prior probability
$\Pr(f\in\calc_{{\rm even}}) = 2/3$, not $1/2$.  It is straightforward 
to check that two classical queries yields no information.  Thus, by 
Theorem~1, a single quantum query is useless.  (It turns out that two 
quantum queries suffice to solve this problem with probability $1$.  
This result and generalizations will be the subject of a subsequent 
publication [\MeyerPommersheimB].)

%--------|---------|---------|---------|---------|---------|---------|
\noindent\Example~3 (P{\eightpoint OLYNOMIAL} 
I{\eightpoint NTERPOLATION}).  Shamir's example of a threshold secret
sharing scheme [\Shamir] provides a distinct family of examples.  Let 
$p$ be prime; let $p-1 > k\in\N$; and let 
$$
\calc = \Bigl\{f:\{1,\ldots,p-1\}\to\Z_p
               \Bigm|
               f(x) = \sum_{i=0}^k a_i x^i\hbox{\ for\ $a_i\in\Z_p$}
        \Bigr\}.
$$
Let $\mu$ be the uniform distribution on $\calc$; this is equivalent 
to choosing each $a_i$ independently and uniformly at random in 
$\Z_p$.  For $j\in\Z_p$, let $\calc_j = \{f\in\calc\mid f(0) = j\}$.
Since the unknown polynomial $f$ has degree $k$, interpolation of the 
$k$ values obtained by $k$ classical queries, {\sl together with any 
value for $f(0)$}, identifies $f$.  Since the value for $f(0)$ is
chosen uniformly at random, this means that any $k$ classical queries
alone give no information about $\calc_j$.  So Theorem~1 tells us that
$\lfloor k/2\rfloor$ quantum queries are useless.  As with 
P{\eightpoint ARITY}, this implies a lower bound for the quantum query 
complexity of P{\eightpoint OLYNOMIAL} I{\eightpoint NTERPOLATION}:

%--------|---------|---------|---------|---------|---------|---------|
\noindent\Theorem\ 2.  {\sl For} P{\eightpoint OLYNOMIAL} 
I{\eightpoint NTERPOLATION}, {\sl $\lfloor k/2\rfloor$ quantum queries 
are useless, and hence the quantum query complexity of } 
P{\eightpoint OLYNOMIAL} I{\eightpoint NTERPOLATION} {\sl is at least 
$\lfloor k/2\rfloor + 1$.}

\medskip
\noindent{\bf 4.  Proof of the main theorem}

%--------|---------|---------|---------|---------|---------|---------|
The proof of Theorem~1 rests upon the following lemma:

%--------|---------|---------|---------|---------|---------|---------|
\noindent\Lemma.  {\sl Let 
$\bigl(\calc,\{\calc_j\mid j\in J\},\mu\bigr)$ be a learning problem.
If $2k$ classical queries are useless, then for any $j$, 
$$
\sum_{f\in\calc_j} \mu(f) \rho_f 
 = 
\mu(\calc_j) \sum_{f\in\calc} \mu(f) \rho_f,
$$
where $\rho_f$ is defined by equation\/} ({\sl 1\/}).

%--------|---------|---------|---------|---------|---------|---------|
\noindent{\sl Proof}.  First note that any matrix 
$B\in\calh\otimes\calh^{\dagger}$ has rows and columns indexed by 
$X\times Y\times Z$.  Since $\calof$ is a permutation matrix, it is 
easy to express the entries of the matrix 
$\calof^{\vphantom\dagger} B \calof^{\dagger}$ in terms of the matrix 
$B$.  If $L=(x,y,z)$ and $M=(u,v,w)$, then
$$
(\calof^{\vphantom\dagger} B \calof^{\dagger})_{L,M} = B_{fL,fM},
                                                              \eqno(2)
$$
where for the triple $L=(x,y,z)$, we define $fL=(x,y+f(x),z)$.  

%--------|---------|---------|---------|---------|---------|---------|
Let $\rho_i$ denote the state after the $i^{\rm th}$ query and after 
applying $U_i$, as in equation (1).  Then
$$
\rho_i 
 = 
U_i^{\vphantom\dagger} 
 \calof^{\vphantom\dagger} 
  \rho_{i-1} 
 \calof^{\dagger} 
U_i^{\dagger},
$$
and from equation (2) and matrix multiplication, it follows that
$$
(\rho_i)_{L,M} 
 = 
\sum_{L',M'} 
 (U_i^{\vphantom\dagger})_{L,L'} 
  (\rho_{i-1})_{fL',fM'}
 (U_i^{\dagger})_{M',M},                                      \eqno(3)
$$
with the sum taken over all $L',M'\in X\times Y\times Z$.  Now apply 
equation (3) iteratively:  

%--------|---------|---------|---------|---------|---------|---------|
First,
$$
(\rho_1)_{L,M} 
 =
\sum_{L_1,M_1} 
 (U_1^{\vphantom\dagger})_{L,L_1} 
  (\rho_{0})_{fL_1,fM_1}
 (U_1^{\dagger})_{M_1,M}.
$$
Note that the quantity being summed depends only on the indices 
$L$, $M$, $L_1$ and $M_1$, and the two function values $f(x_1)$ and 
$f(u_1)$, where $x_1$ and $u_1$ are the first coordinates of $L_1$ and 
$M_1$, respectively.  (It also depends on $\rho_0$ and the unitary 
matrix $U_0$, but these are fixed.)

%--------|---------|---------|---------|---------|---------|---------|
Second,
$$
(\rho_2)_{L,M} 
 = 
\sum_{L_1,M_1,L_2,M_2}
 (U_2^{\vphantom\dagger})_{L,L_2} 
  (U_1^{\vphantom\dagger})_{fL_2,L_1} 
   (\rho_{0})_{fL_1,fM_1}
  (U_1^{\dagger})_{M_1,fM_2}
 (U_2^{\dagger})_{M_2,M}.
$$
Here the quantity being summed depends on the indices $L$, $M$, $L_1$,
$L_2$, $M_1$ and $M_2$, and the four function values $f(x_1)$, 
$f(x_2)$, $f(u_1)$ and $f(u_2)$.

%--------|---------|---------|---------|---------|---------|---------|
Continuing in this manner, the final density matrix after $k$ queries, 
$\rho_k=\rho_f$, is given by
$$
\rho_f 
 = 
\sum_I Q_I\bigl(f(x_1),\ldots,f(x_k),f(u_1),\ldots,f(u_k)\bigr),
$$
where the sum is taken over all tuples 
$I = (L_1,\ldots,L_k,M_1,\ldots,M_k)\in(X\times Y\times Z)^{2k}$, and 
$Q_I(f(x_1),\ldots,f(x_k),f(u_1),\ldots,f(u_k))
 \in\calh\otimes\calh^{\dagger}$ is a matrix that depends only on the 
index $I$ and the $2k$ function values shown.

%--------|---------|---------|---------|---------|---------|---------|
Thus, for any $j\in J$,
$$
\sum_{f\in\calc_j} \mu(f) \rho_f 
 =
\sum_I
 \sum_{f\in\calc_j} 
  \mu(f)Q_I\bigl(f(x_1),\ldots,f(x_k),f(u_1),\ldots,f(u_k)\bigr).
                                                              \eqno(4)
$$
Regrouping, the right hand side of equation (4) becomes
$$
\sum_I 
 \sum_{\{y_i\},\{v_i\}} 
  \mu\bigl(f\in\calc_j 
           \hbox{\ and\ } 
           f(x_i) = y_i, f(u_i) = v_i, i\in\{1,\ldots,k\} 
     \bigr)
   Q_I(y_1,\ldots,y_k,v_1,\ldots,v_k),
$$
with the inner sum taken over all 
$y_1,\ldots,y_k,v_1,\ldots,v_k\in Y$.  But by the hypothesis that $2k$ 
classical queries yield no information, 
$$
\eqalign{
&\mu\bigl(f\in\calc_j 
         \hbox{\ and\ } 
         f(x_i) = y_i,f(u_i) = v_i, i\in\{1,\ldots,k\}
   \bigr)                                                          \cr
&\qquad\qquad\qquad= 
\mu(\calc_j)
\mu\bigl(f(x_i) = y_i,f(u_i) = v_i, i\in\{1,\ldots,k\} 
   \bigr).                                                         \cr
}
$$
Thus equation (4) becomes
$$
\eqalign{
&\sum_{f\in\calc_j} \mu(f) \rho_f                                  \cr
&\qquad=
\mu(\calc_j) 
\sum_{I,\{y_i\},\{v_i\}} 
 \mu\bigl(f(x_i) = y_i,f(u_i) = v_i, i\in\{1,\ldots,k\} 
    \bigr) 
 Q_I(y_1,\ldots,y_k,v_1,\ldots,v_k).                               \cr
}                                                             \eqno(5)
$$
Summing equation (5) over all $j$ gives
$$
\sum_{f\in\calc} \mu(f) \rho_f  
 = 
\sum_{I,\{y_i\},\{v_i\}} 
 \mu\bigl(f(x_i) = y_i,f(u_i) = v_i, i\in\{1,\ldots,k\} 
    \bigr) 
 Q_I(y_1,\ldots,y_k,v_1,\ldots,v_k),
$$
whence the the lemma follows.                         \hfill\endproof

%--------|---------|---------|---------|---------|---------|---------|
\noindent{\sl Proof of Theorem 1}.  The statement of the theorem is 
that the probability of $f$ being in $\calc_j$ does not change if $s$ 
is observed after $k$ queries, \ie, for any $j\in J$ and $s\in S$, we 
need to show that
$$
\sum_{f\in \calc_j} \mu(f\mid s) = \mu(\calc_j).
$$
To prove this, calculate the probability of $f$ having been the chosen 
function conditioned on having observed $s$, using Bayes' Theorem:
$$
\mu(f\mid s) 
 = 
{\Tr(\rho_f\Pi_s)\mu(f)
 \over\sum_{g\in\calc}
 \Tr(\rho_g\Pi_s)\mu(g)
}
$$
Thus,
$$
\sum_{f\in\calc_j} \mu(f\mid s) 
 = 
{\Tr\Bigl(\bigl(\sum_{f\in\calc_j} \mu(f)\rho_f   
          \bigr) 
          \Pi_s
    \Bigr)   
 \over  
 \Tr\Bigl(\bigl(\sum_{g\in\calc} \mu(g)\rho_g
          \bigr) 
          \Pi_s
    \Bigr)   
}.                                                            \eqno(6)
$$
Applying the Lemma, the quotient on the right hand side of equation 
(6) reduces to $\mu(\calc_j)$, establishing the theorem.
                                                       \hfill\endproof

\medskip
\noindent{\bf 5.  Conclusion}

\nobreak
%--------|---------|---------|---------|---------|---------|---------|
As we noted in the introduction, Theorem~1 implies a lower bound on 
the quantum query complexity of certain learning problems:

%--------|---------|---------|---------|---------|---------|---------|
\noindent\Theorem\ 3.  {\sl Let 
$\bigl(\calc,\{\calc_j\mid j\in J\},\mu\bigr)$ be a learning problem.
Suppose that $2k$ classical queries are useless.  Then the quantum
query complexity of the problem is at least $k+1$.}

%--------|---------|---------|---------|---------|---------|---------|
The uselessness of some number of quantum queries in learning 
problems with two subclasses also has a consequence for {\sl amplified 
impatient learning\/} [\HMPPR]:  If in addition to the {\sl membership 
oracle\/} (the oracle that returns function values), we have access to 
an {\sl equivalence oracle\/} (an oracle that answers the questions of 
the form ``Is $f\in\calc_j$?''), a commonly studied situation in 
computational learning theory [\Angluin], we can implement 
{\sl amplitude amplification\/} 
[\Grover,\BrassardHoyer,\GroverB,\BHT,\BHMT] after any number of 
quantum queries.  If $k$ quantum queries to the membership oracle are 
useless, however, amplitude amplification works exactly as well if it 
is implemented immediately, \ie, after no quantum queries, as when it 
is implemented after $k$ or fewer quantum queries.

%--------|---------|---------|---------|---------|---------|---------|
These results encourage further investigation of the quantum query 
complexity of, and quantum algorithms for, learning problems in which
some number of classical queries are useless.  These include problems
in the families exemplified by Examples~2 and 3.  We will address some 
of these questions in a forthcoming paper [\MeyerPommersheimB].

\medskip
\noindent{\bf Acknowledgements}

\nobreak
%--------|---------|---------|---------|---------|---------|---------|
This work has been partially supported by the National Science 
Foundation under grant ECS-0202087 and by the Defense Advanced 
Research Projects Agency as part of the Quantum Entanglement Science
and Technology program under grant N66001-09-1-2025.

\medskip
\global\setbox1=\hbox{[00]\enspace}
\parindent=\wd1
\noindent{\bf References}

\nobreak
\item{[\Simon]}
%--------|---------|---------|---------|---------|---------|---------|
\simon,
``On the power of quantum computation'',
in S. Goldwasser, ed.,
{\sl Proceedings of the 35th Annual Symposium on Foundations of 
     Computer Science}, Santa Fe, NM, 20--22 November 1994
(Los Alamitos, CA:  IEEE 1994) 116--123;\hfb
\simon,
``On the power of quantum computation'',
\SIAMJC\ {\bf  26} (1997) 1474--1483.

\parskip=0pt
\item{[\Shor]}
%--------|---------|---------|---------|---------|---------|---------|
\shor,
``Algorithms for quantum computation:  discrete logarithms and 
  factoring'',
in S. Goldwasser, ed.,
{\sl Proceedings of the 35th Symposium on Foundations of Computer 
Science}, Santa Fe, NM, 20--22 November 1994
(Los Alamitos, CA:  IEEE Computer Society Press 1994) 124--134;\hfb
\shor,
``Polynomial-time algorithms for prime factorization and discrete 
  logarithms on a quantum computer'',
\SIAMJC\ {\bf 26} (1997) 1484--1509.

\item{[\BonehLipton]}
%--------|---------|---------|---------|---------|---------|---------|
D. Boneh and R. J. Lipton,
``Quantum cryptanalysis of hidden linear forms'',
in D. Coppersmith, ed.,
{\sl Proceedings of Crypto '95}, 
{\sl Lecture Notes in Computer Science\/} {\bf 963} 
(Berlin:  Springer-Verlag 1995) 424--437.

\item{[\Jozsa]}
%--------|---------|---------|---------|---------|---------|---------|
R. Jozsa, 
``Quantum algorithms and the Fourier transform'',
\PRSLA\ {\bf 454} (1998) 323--337.

\item{[\Angluinsurvey]}
%--------|---------|---------|---------|---------|---------|---------|
See, \eg,
D. Angluin,
``Computational learning theory:  Survey and selected bibliography'',
in
{\sl Proceedings of the Twenty-Fourth Annual ACM Symposium on Theory
     of Computing\/}
(New York:  ACM 1992) 351--369.

\item{[\Grover]}
%--------|---------|---------|---------|---------|---------|---------|
\grover,
``A fast quantum mechanical algorithm for database search'',
in {\sl Proceedings of the Twenty-Eighth Annual ACM Symposium on 
  the Theory of Computing},
Philadelphia, PA, 22--24 May 1996 
(New York:  ACM 1996) 212--219;\hfb
\grover, 
``Quantum mechanics helps in searching for a needle in a haystack'', 
%{\tt quant-ph/9706033};
\PRL\  {\bf 79} (1997) 325--328.

\item{[\Deutsch]}
%--------|---------|---------|---------|---------|---------|---------|
D. Deutsch,
``Quantum theory, the Church-Turing principle and the universal 
  quantum computer'',
\PRSLA\ {\bf 400} (1985) 97--117.

\item{[\CEMM]}
%--------|---------|---------|---------|---------|---------|---------|
R. Cleve, A. Ekert, C. Macchiavello and M. Mosca,
``Quantum algorithms revisited'',
\PRSLA\ {\bf 454} (1998) 339--354.

\item{[\BBCMdW]}
%--------|---------|---------|---------|---------|---------|---------|
R. Beals, H. Buhrman, R. Cleve, M. Mosca and R. de Wolf,
``Quantum lower bounds by polynomials'',
\JACM\ {\bf 48} (2001) 778--797.

\item{[\FGGS]}
%--------|---------|---------|---------|---------|---------|---------|
E. Farhi, J. Goldstone, S. Gutmann and M. Sipser,
``Limit on the speed of quantum computation in determining parity'',
\PRL\ {\bf 81} (1998) 5442--5444.

\item{[\Shamir]}
%--------|---------|---------|---------|---------|---------|---------|
A. Shamir,
``How to share a secret'',
\CACM\ {\bf 22} (1979) 612--613.

\item{[\BVdW]}
%--------|---------|---------|---------|---------|---------|---------|
H. Buhrman, N. Vereshchagin and R. de Wolf, 
``On computation and communication with small bias'',
{\sl Proceedings of the 22nd Annual Conference on Computational 
     Complexity}, 
San Diego, CA, 13-16 June 2007
(Los Alamitos, CA:  IEEE 2007) 24--32.

\item{[\MeyerPommersheim]}
%--------|---------|---------|---------|---------|---------|---------|
\dj,
``Single query learning from abelian and non-abeli\-an Hamming 
  distance oracles'',
{\tt arXiv:0912.0583v1 [quant-ph]}.

\item{[\MNR]}
%--------|---------|---------|---------|---------|---------|---------|
A. Montanaro, H. Nishimura and R. Raymond,
``Unbounded error quantum query complexity'',
%{\tt arXiv:0712.1446v1 [quant-ph]};
in
S.-H. Hong, N. Nagamochi and T. Fukunaga, eds.,
{\sl Algorithms and Computation}, 
proceedings of the 19th International Symposium, ISAAC 2008, 
Gold Coast, Australia, 15-17 December 2008,
{\sl Lecture Notes in Computer Science\/} {\bf 5369}
(Berlin:  Springer-Verlag 2008) 919--930.

\item{[\MeyerPommersheimB]}
%--------|---------|---------|---------|---------|---------|---------|
\dj,
``Multi-query quantum algorithms for summation'',
in preparation.

\item{[\HMPPR]}
%--------|---------|---------|---------|---------|---------|---------|
M. Hunziker, \dajm, J. Park, J. Pommersheim and M. Rothstein,
``The geometry of quantum learning'',
%{\tt quant-ph/0309059};
\QIP\ DOI: 10.1007/s11128-009-0129-6 (2009) 1--21.

\item{[\Angluin]}
%--------|---------|---------|---------|---------|---------|---------|
D. Angluin,
``Queries and concept learning'',
\ML\ {\bf 2} (1988) 319--342.

\item{[\BrassardHoyer]}
%--------|---------|---------|---------|---------|---------|---------|
G. Brassard and P. H{\o}yer,
``An exact quantum polynomial-time algorithm for Simon's problem'',
%{\tt quant-ph/9704027};
{\sl Proceedings of 5th Israeli Symposium on Theory of
      Computing and Systems}, Ramat-Gan, Israel 17--19 June 1997
(Los Alamitos, CA:  IEEE 1997) 12--23.

\item{[\GroverB]}
%--------|---------|---------|---------|---------|---------|---------|
\grover,
``A framework for fast quantum mechanical algorithms'',
%{\tt quant-ph/ 9711043};
in
{\sl Proceedings of the 30th Annual ACM Symposium on Theory of 
     Computing},
Dallas, TX, 23--26 May 1998
(New York:  ACM 1998) 53--62.

\item{[\BHT]}
%--------|---------|---------|---------|---------|---------|---------|
G. Brassard, P. H{\o}yer and A. Tapp,
``Quantum counting'',
%{\tt quant-ph/9805082};
{\sl Proceedings of the 25th International Colloquium on Automata,
Languages, and Programming}, {\AA}lborg, Denmark, 13--17 July 1998,
{\sl Lecture Notes in Computer Science\/} {\bf 1443}
(Berlin:  Springer-Verlag 1998) 820--831.

\item{[\BHMT]}
%--------|---------|---------|---------|---------|---------|---------|
G. Brassard, P. H{\o}yer, M. Mosca and A. Tapp,
``Quantum amplitude amplification and estimation'',
%{\tt quant-ph/0005055};
in
S. J. Lomonaco, Jr.\ and H. E. Brandt, eds.,
{\sl Quantum Computation and Information},
{\sl Contemporary Mathematics\/} {\bf 305}
(Providence, RI:  AMS 2002) 53--74.

\medskip
\parindent=0pt\parskip=10pt
\noindent{\bf Appendix}

%--------|---------|---------|---------|---------|---------|---------|
In this appendix, we focus on the special case in which we are trying 
to compute a Boolean function, or partial Boolean function, of $N$-bit
strings.  That is, we assume that (a) the concept class $\calc$ 
consists of Boolean-valued functions, \ie, the codomain is 
$Y=\{0,1\}$; and (b) the partition of $\calc$ has exactly two parts 
$\calc=\calc_0\sqcup\calc_1$.  We show that the polynomial method of 
[\BBCMdW], together with an observation of [\BVdW], can be used to 
prove Theorem~1 for this special case.  Similar ideas appear in 
Section~3 of [\MNR].  As noted in the Introduction, it seems that 
these ideas cannot be used to prove Theorem~1, which is not limited 
by restrictions (a) or (b).

%--------|---------|---------|---------|---------|---------|---------|
We base our proof of this special case of Theorem~1 on the following 
result, which gives a general relation between $k$-query quantum 
algorithms and $2k$-query classical algorithms. 

%--------|---------|---------|---------|---------|---------|---------|
\noindent\Theorem\ 4.  {\sl Suppose 
$\bigl(\calc,\{\calc_j\mid j\in J\},\mu\bigr)$ is a learning problem, 
as described above, such that $Y=\{0,1\}$ and $J=\{0,1\}$.  Given a 
$k$-query quantum algorithm, for each $N$-bit string $f$ of function
values computed by the oracle, denote by $p(f)$ the probability that 
the quantum algorithm outputs $0$.  Then there exists a positive real 
number $T$ and a $2k$-query \hbox{\rm (}randomized\/\hbox{\rm )} 
classical algorithm whose output probability for $f$ is given by
$$
p_{{\rm classical}}(f) 
 = {1 \over T}\Bigl(p(f)-{1 \over 2}\Bigr) + {1 \over 2}.
$$
That is, for each $f$, the bias of the classical algorithm away from 
${1\over2}$ is $T^{-1}$ times the bias of the quantum algorithm away 
from ${1\over 2}$.}

%--------|---------|---------|---------|---------|---------|---------|
Note that this theorem does not require the existence of a prior 
distribution on $\calc$.

%--------|---------|---------|---------|---------|---------|---------|
\noindent{\sl Proof}.  For convenience, we will assume that 
$X=\{1,\ldots,N\}$, and we will identify $Y^X$ with $N$-bit strings 
$f=\bigl(f(1),\ldots,f(N)\bigr)$.  $\calc$ is then a subset of $N$-bit 
strings.  Suppose we are given any $k$-query quantum algorithm.  Then 
the arguments of [\BBCMdW] show that there exists a squarefree 
polynomial $p(f)$ of degree at most $2k$ with real coefficients such 
that for any $f\in\calc$ evaluated by the oracle, $p(f)$ is the 
probability that the quantum algorithm outputs $0$.

%--------|---------|---------|---------|---------|---------|---------|
We now change variables, so as to identify functions from 
$\{0,1\}^N\rightarrow\{0,1\}$ with functions 
$\{-1,1\}^N\rightarrow\{-1,1\}$.  Specifically, we introduce the 
polynomial 
$$
q(w_1,\ldots,w_N) 
= 2p\biggl({{w_1+1}\over 2},\ldots,{{w_N+1}\over{2}}\biggr)-1.
$$
Then $q(w)$ is a squarefree polynomial of degree at most $2k$ with 
real coefficients, and has the property that for any $w\in\{-1,1\}^N$, 
the probability that the corresponding $f\in\{0,1\}^N$ leads to an 
output of $0$ is equal to $p(x)=\bigl(1+q(w)\bigr)/2$.  Then we have
$$
q(w) = \sum_{S} \qhat(S) w_S,
$$
where the sum is over all subsets $S$ of $\{1,\ldots,N\}$ of size less 
than or equal to $2k$, and $w_S$ denotes the product of $w_i$ with 
$i\in S$.  Let $T = \sum_S |\qhat(S)|$.

%--------|---------|---------|---------|---------|---------|---------|
We now introduce a classical algorithm, following the observation of 
Buhrman, \etal~[\BVdW].  First note that the absolute value of 
$\qhat/T$ defines a probability distribution on the subsets $S$ of 
$\{1,\ldots,N\}$ of size less than or equal to $2k$.  Begin by picking 
a random subset $S$ according to this distribution.  By invoking the 
classical oracle at most $2k$ times, compute $w_S$.  Then according to 
whether $\sign\bigl(\qhat(S)\bigr)w_S$ is $1$ or $-1$, output $0$ or 
$1$, respectively.

%--------|---------|---------|---------|---------|---------|---------|
We claim that for any $f\in \calc\subseteq \{0,1\}^N$, the probability 
that this classical algorithm outputs $0$ equals 
$p_{{\rm classical}}(f) 
 = \bigl(p(f)-{1\over 2}\bigr)/T + {1\over 2}$.  To see this, note 
that the probability of outputting $0$ is 
$$
\sum_S {|\qhat(S)| \over T} \delta_S,
$$
where $\delta_S=1$ if $\sign\bigl(\qhat(S)\bigr)w_S = 1$, and $0$ 
otherwise.  This simplifies to 
$$
\sum_S {|\qhat(S)|\over T} \biggl({{\sign(\qhat(S))w_S + 1}\over{2}}
                           \biggr) 
= {1\over2T}\Bigl(T+\sum_S \qhat(S) w_S 
            \Bigr) 
= {T+q(w)\over 2T} 
= {1\over T}\Bigl(p(f)-{1\over 2}\Bigr) + {1\over 2},
$$
as desired.                                            \hfill\endproof

%--------|---------|---------|---------|---------|---------|---------|
As a consequence we have the following special case of Theorem~1:

%--------|---------|---------|---------|---------|---------|---------|
\noindent\Corollary\ 5.  {\sl For any learning problem 
$\bigl(\calc,\{\calc_j\mid j\in J\},\mu\bigr)$ 
as described above, with $Y=\{0,1\}$ and $J=\{0,1\}$, if $2k$ 
classical queries are useless, then $k$ quantum queries are useless.}

%--------|---------|---------|---------|---------|---------|---------|
\noindent{\sl Proof}.  Suppose that $2k$ classical queries are 
useless.  Given any $k$-query quantum algorithm, consider the 
corresponding $2k$-query classical algorithm given by Theorem~4.  
Since this algorithm is useless, we have
$$
{
 \sum_{f\in\calc_0} \mu(f)\bigl(p(f) - {1\over2}\bigr)/T
 \over
 \sum_{f\in\calc} \mu(f) \bigl(p(f) - {1\over2}\bigr)/T 
}
=
\sum_{f\in\calc_0} \mu(x).
$$
It follows that
$$
{
 { \sum_{f\in\calc_0} \mu(f) p(f) }
\over
{ \sum_{f\in\calc} \mu(f) p(f) }
}
=
\sum_{f\in\calc_0} \mu(f).
$$
In other words, the quantum algorithm is also useless.
                                                     \hfill\endproof

\bye

\noindent{\bf $\pi$. Other stuff to possibly incorporate somewhere:}

proof by triple slit.  Dave had the idea to make this a separate physicsy paper (2 columns).

proof by Helstrom (some of this is in the morgue)

No info implies amplitude amplification will not work as a multiple quantum query algorithm

Anything else we want to add about forthcoming $\Z_N\rightarrow\Z_k$ work.  ( Eg. A single quantum query does no better than a single classical query 
for the $\Z_3\to\Z_3$ mod 3 sum problem.   Two quantum queries solves the $\Z_3\to\Z_3$ 3-parity problem with probability 1. One classical and one quantum query solves the $\Z_3\to\Z_3$ even/odd 
problem with probability 5/6. $k-1$ quantum queries solves the $\Z_k\to\Z_k$ $k$-parity problem with probability 1.  Van Dam stuff, etc.  )

\vfill\eject

\noindent{\bf $\infty$. Morgue}

Let $\psi\in C^X\otimes \C^Y$ denote the initial state.  Then after the oracle call, the system is in the state $\calof \psi$, and has a density matrix given by
$$
\rho_f = | \calof \psi \rangle \langle \calof \psi |.
$$

We wish to show that for an arbitrary $\psi$ and an arbitrary POVM $\{X_u\}$, the success probability, which is given by
$$
\Pr({\rm success})= \sum_u \sum_{f\in \calc_u}  \Pr(f) \Tr(\rho_f X_u),
$$
does not exceed $\Pr(\calc_{\uhat})$.    For this purpose, we rely on the results of [Holevo], which gives a criterion for the optimum measurement to use to distinguish quantum states.  (See also the book of [Helstrom].)    We now summarize this result in the Bayesian case, we will apply to our problem.

For each $u$, define
$$
F_u = - \sum_{f\in \calc_u} \Pr(f)\rho_f, 
$$
and define
$$
\Lambda = -\sum_u \sum_{f\in \calc_u} \Pr(f) \rho_f X_u.
$$
Then for the optimality of the measurement $\{X_u\}$, it is sufficient to show that $\Lambda$ is Hermitian, and that for all $u$,
$$
F_u \geq \Lambda.
$$

We claim that for any choice of initial state $\psi$, the measurement given by $X_{\uhat} = I $ and $X_u = 0$ for all $u\neq \uhat$ satisfies these two conditions, and is therefore optimal.  Note that the success probability for this measurement is simply $\Pr(\calc_{\uhat})$, and therefore the theorem follows.

The first condition, that $\Lambda$ is Hermitian, follows easily, since $\Lambda = F_{\uhat}$.   To verify the second condition,  we use the following lemma.

\noindent\Lemma\ 2.  For any $u$, the operator ${1 \over {\Pr(\calc_u})} F_u$ is independent of $u$.

From the lemma, it follows that for any $u$,
$$
F_u -\Lambda = \biggl(  {{\Pr(\calc_u})\over{\Pr(\calc_{\uhat})}}  -1  \biggr) F_{\uhat},
$$
and hence $F_u -\Lambda \geq 0$, so the second condition is verified, and we are done.

\noindent{\sl Proof of Lemma 2}.

Let $x,w \in X$ and $s,t  \in Y$.  Then 
$$
(\rho_f)_{x,s,w,t} = \bar{\psi}_{x,s-f(x)} \psi_{w,t-f(w)}.
$$
Adding over $f\in\calc_u$, we find
$$
\biggl(   \sum_{f\in\calc_u}  \Pr(f) \rho_f \biggr)_{x,s,w,t} = \sum_{a,b\in Y} \sum_{f} \Pr(f) \bar{\psi}_{x,s-a} \psi_{w,t-b},
$$
where the inner sum is taken over those $f\in\calc_u$ such that $f(x)=a$ and $f(w) = b$.  Thus
$$
\biggl(   \sum_{f\in\calc_u}  \Pr(f) \rho_f \biggr)_{x,s,w,t} =\sum_{a,b\in Y} \Pr\biggl(f\in \calc_u {\rm \ and \ } f(x) = a {\rm\ and \ } f(w) = b\biggr)\  \bar{\psi}_{x,s-a} \psi_{w,t-b}
$$.

But by the hypothesis that two classical queries do not change the probability of $f$ being in $\calc_u$, we have
$$
\Pr\biggl(f\in \calc_u {\rm\ and \ } f(x) = a {\rm\ and \  }f(w) = b\biggr) =  \Pr(\calc_u)\cdot  \Pr\biggl(f(x) = a {\rm\ and\ } f(w) = b\biggr).
$$
Thus we see that
$$
\biggl(   \sum_{f\in\calc_u} \Pr(f) \rho_f   \biggr)_{x,s,w,t} =\Pr(\calc_U) \sum_{a,b\in Y} \Pr\biggl( f(x) = a {\rm\ and\ } f(w) = b\biggr) \ \bar{\psi}_{x,s-a} \psi_{w,t-b}.
$$
The sum on the right hand side is independent of $u$, and the lemma follows. \hfill\endproof

\noindent\Lemma\ 4.  For any $u$, 
$$
\sum_{f\in\calc_u} \Pr(f) \rho_f  = \Pr(\calc_u) \sum_{f\in\calc} \Pr(f) \rho_f.
$$

\noindent{\sl Proof of Lemma 4}.

The proof is almost identical to that of Lemma 2.  The right hand side of the last formula in that proof is easily seen to equal  
$$
 \Pr(\calc_u) \sum_{f\in\calc} \Pr(f) \rho_f.
 $$
 \hfill\endproof

With no queries to the oracle, our best strategy is to chose the class $\calc_{\uhat}$ which contains the function with the greatest probability.  

As before let $\psi$ be the initial state.  This time, we measure $\calof\psi$ according to some POVM $\{\Pi_k\}$ (not indexed by $u$, but by some other arbitrary set), and ask what we can conclude about $Pr(f|k)$, the probability that the chosen function is $f$ given that the result of the measurement is $k$

\noindent\Theorem\ 1.  Suppose that using two classical queries does not change the probability of $f$ being in $\calc_u$ for any $u$.   Then a single quantum query yields no improvement in success probability.  

\noindent{\sl Proof}.

\bye